\documentclass[letterpaper,10pt,twocolumn,twoside,final,journal]{ieeetj}

\usepackage[utf8]{inputenc}
\usepackage[T1]{fontenc}
\usepackage{cite}
\usepackage{amsmath,amssymb,amsfonts}
\usepackage{algorithmic}
\usepackage{graphicx,color}
\usepackage{textcomp}
\usepackage{xcolor}
\usepackage{algorithm,algorithmic}
\usepackage{booktabs}
\usepackage{nicefrac}
\usepackage{tikz}
\usepackage{tikzsymbols}
\usepackage{etoolbox}
\usepackage{url}
\usepackage{newtxtext}
\usepackage{newtxmath}
\usepackage{inconsolata}
\usepackage{microtype}
\usepackage{framed}
\usepackage{orcidlink}
\usepackage{verbatim}
\usepackage{csquotes}
\usepackage[english]{babel}

\usepackage{hyperref}
\hypersetup{
    colorlinks=true,
    linkcolor=blue,
    citecolor=blue,
    urlcolor=black,
}

\graphicspath{{figs/}}
\makeatletter
\def\input@path{{figs/}}
\def\fnum@figure{\MakeUppercase{\figurename~\thefigure}}
\def\fnum@table{\MakeUppercase{\tablename}~\thetable}
\renewcommand{\thetable}{\Roman{table}} 
\makeatother
 
\newrobustcmd{\code}[1]{{\small\ttfamily #1}}
\newrobustcmd{\class}[1]{{\small\ttfamily \textquoteleft #1\textquoteright}}

\newenvironment{terminal}
  {\par\medskip\small\ttfamily\verbatim}
  {\endverbatim\medskip}

\definecolor{lightgray}{gray}{0.9}
\newcommand{\info}[1]{%
  \begingroup
  \small\ttfamily
  \setlength{\tabcolsep}{4pt}
  \setlength{\fboxsep}{0pt}
  \noindent\colorbox{lightgray}{\parbox{\dimexpr\linewidth-2\fboxsep\relax}{\strut#1}}%
  \endgroup
}

\def\BibTeX{{\rm B\kern-.05em{\sc i\kern-.025em b}\kern-.08em
    T\kern-.1667em\lower.7ex\hbox{E}\kern-.125emX}}
\AtBeginDocument{\definecolor{tmlcncolor}{cmyk}{0.93,0.59,0.15,0.02}\definecolor{NavyBlue}{RGB}{0,86,125}}
 
\def\OJlogo{}
\def\seclogo{}

\def\authorrefmark#1{\ensuremath{^{\textbf{#1}}}}

\begin{document}
\receiveddate{2 September, 2024}
\reviseddate{26 November, 2024}
\accepteddate{24 April, 2025}
\currentdate{29 April, 2025}
\doiinfo{10.1109/TP.2025.3566052}
\jvol{2}
\pubyear{2025}

\markboth{A MULTI-LANGUAGE TOOLKIT FOR THE SEMI-AUTOMATED CHECKING OF RESEARCH OUTPUTS}{PREEN ET AL.}

\title{A Multi-Language Toolkit for the Semi-Automated Checking of Research Outputs}

\author{RICHARD J. PREEN~\orcidlink{0000-0003-3351-8132}\authorrefmark{1}, MAHA ALBASHIR\authorrefmark{1}, SIMON DAVY~\orcidlink{0000-0001-9890-3619}\authorrefmark{2}, AND JIM SMITH~\orcidlink{0000-0001-7908-1859}\authorrefmark{1}}
\affil{Department of Computer Science and Creative Technologies, University of the West of England, UK}
\affil{Bennett Institute for Applied Data Science, University of Oxford, UK}
\corresp{CORRESPONDING AUTHOR\@: RICHARD J. PREEN (email: richard2.preen@uwe.ac.uk).}
\authornote{This work was funded by UK Research and Innovation under Grant Number MC\_PC\_23006 as part of Phase 1 of the Data and Analytics Research Environments UK programme (DARE UK), delivered in partnership with Health Data Research UK (HDR UK) and Administrative Data Research UK (ADR UK). The specific project was `Semi-Automated Checking of Research Outputs' (SACRO). It has also been supported by the University of the West of England, as part of a wider project funded under the Expanding Research Excellence scheme. No new data were created during this research.}

\begin{abstract}
This article presents a free and open source toolkit that supports the semi-automated checking of research outputs (SACRO) for privacy disclosure within secure data environments. SACRO is a framework that applies best-practice principles-based statistical disclosure control (SDC) techniques on-the-fly as researchers conduct their analyses. SACRO is designed to assist human checkers rather than seeking to replace them as with current automated rules-based approaches. The toolkit is composed of a lightweight Python package that sits over well-known analysis tools that produce outputs such as tables, plots, and statistical models. This package adds functionality to (i) automatically identify potentially disclosive outputs against a range of commonly used disclosure tests; (ii) apply optional disclosure mitigation strategies as requested; (iii) report reasons for applying SDC\@; and (iv) produce simple summary documents trusted research environment staff can use to streamline their workflow and maintain auditable records. This creates an explicit change in the dynamics so that SDC is something done with researchers rather than to them, and enables more efficient communication with checkers. A graphical user interface supports human checkers by displaying the requested output and results of the checks in an immediately accessible format, highlighting identified issues, potential mitigation options, and tracking decisions made. The major analytical programming languages used by researchers (Python, R, and Stata) are supported by providing front-end packages that interface with the core Python back-end. Source code, packages, and documentation are available under MIT license at \url{https://github.com/AI-SDC/ACRO}
\end{abstract}


\maketitle

\section{INTRODUCTION}
Statistical agencies and other custodians of secure data facilities such as trusted research environments (TREs)~\cite{Hubbard:2020} provide researchers with access to confidential data under the `Five Safes' data governance framework~\cite{Green:2023}. This enforces five orthogonal layers of safety procedures with the last requiring the explicit checking of research outputs for disclosure risk. This can be a time-consuming and costly task, requiring skilled staff, and is currently impeding efforts to scale-up the use of TREs~\cite{DARE:2022}.

Existing solutions have attempted to automate output statistical disclosure control (SDC) with rules-based approaches. However, defining hard rules that should apply regardless of context is extremely difficult~\cite{Ritchie:2015} and fully-automated solutions are unable to address a number of potential vulnerabilities such as secondary disclosure through differencing attacks.

This article discusses the development of a free and open source toolkit for semi-automating the SDC of routine research outputs such as tables, plots, and statistical models (SACRO). SACRO implements a principles-based SDC methodology, which aims to assist rather than replace human checkers, and ensure that researchers have met minimum required standards before submitting their request for release. The goal is to make the clearance process more efficient and timely, and to allow the skilled checkers to focus their attention on the less straightforward cases.

The SACRO toolkit is composed of (i) a Python package (ACRO) that implements a number of automated checks and optional mitigation strategies that researchers use on-the-fly as they conduct their analyses; and (ii) a graphical user interface (GUI) that uses the reports generated with ACRO to assist checkers in reviewing outputs, tracking decisions, and providing an auditable record (SACRO viewer). Additional languages such as R and Stata are supported with wrapper packages that interface with the Python ACRO back-end.

More specifically, the ACRO tool assists researchers and output checkers by distinguishing between research output that requires further analysis, and output that cannot be published because of a substantial risk of disclosing private data. The tool produces summaries that indicate whether the automated tests passed, failed, or require review. A status of ``pass'' does not indicate that the tool has certified that an output is safe for publication, but that it has passed a minimum level of checks that staff normally would have had to manually perform and provides information to help expedite further review.

The details of the checks and any mitigations are summarised for skilled checkers to assist in their review. This is achieved through the use of Python's capacity to override standard commands for creating tables, regressions, and other queries. This keeps the syntax identical while simultaneously augmenting analysis commands by running and reporting appropriate disclosure risk assessments. A schematic illustration of the SACRO workflow is shown in Figure~\ref{fig:schematic}.

\begin{figure*}[t]
    \centering
    \resizebox{\textwidth}{8cm}{%
    \begin{tikzpicture}[font=\sffamily]
        \setlength{\leftmargini}{1.3em}
        \definecolor{green}{HTML}{c0edc0}
        \definecolor{blue}{HTML}{A7C7E7}
        \node(researcher) at (0,5.0)[align=center] {\Strichmaxerl[7]\\Researcher};
        \node(checker) at (0,0.0)[align=center] {\Strichmaxerl[7]\\TRE Staff};
        \node(frontend) at (7,5)[align=center, rectangle, rounded corners=5pt, draw, fill=green, thick, minimum width=25mm, minimum height=60mm] {};
        \node(functions) at (7,7)[align=center, text width=25mm] {{\bfseries Light-Weight Translation Functions}};
        \node(python) at (7,5.5)[align=center, rectangle, draw, fill=white, thick, minimum width=15mm, minimum height=5mm] {Python};
        \node(r) at (7,4.5)[align=center, rectangle, draw, fill=white, thick, minimum width=15mm, minimum height=5mm] {R};
        \node(stata) at (7,3.5)[align=center, rectangle, draw, fill=white, thick, minimum width=15mm, minimum height=5mm] {Stata};
        \node(other) at (7,2.5)[align=center, rectangle, draw, fill=white, thick, minimum width=15mm, minimum height=5mm] {\ldots};
        \node(output) at (9,0)[align=center, text width=65mm, rectangle, draw, thick] {Excel spreadsheet or JSON file with details and recommendations for each requested output};
        \node(risk) at (15,0)[align=center, text width=30mm, rectangle, draw, thick] {TRE-specific file detailing risk appetite};
        \node(backend) at (11,5)[align=center, rectangle, rounded corners=5pt, draw, thick, fill=green, text width=25mm, minimum height=60mm] {%
            {\bfseries Disclosure Control Checks (Python)}\\
            \vspace{2mm}
            Tests:\\
                - threshold\\
                - dominance\\
                - degrees of freedom\\
            \vspace{2mm}
            Applies:\\
                - cell suppression\\
                - others
        };
        \node(libraries) at (15,5)[align=center, fill=blue, text width=25mm, minimum height=60mm, rectangle, draw, thick] {Standard Python Libraries:\\ \vspace{5mm} {\bfseries Pandas} for tables\\  \vspace{5mm}{\bfseries statsmodels} for regression};
        \draw [black, solid, line width=2pt, <->, >=stealth]
        (backend) edge node [above] {} (libraries);
        \draw (researcher.30) edge[black, solid, line width=2pt, ->, >=stealth] node[align=center, text width=35mm, yshift=5mm] {Analysis commands prefixed by acro} (frontend.156);
        \draw (frontend.195) edge[black, solid, line width=2pt, ->, >=stealth] node[yshift=2mm] {SDC output} (researcher.-20);
        \draw (frontend.30) edge[black, solid, line width=2pt, ->, >=stealth] node {} (backend.152);
        \draw [black, dashed, line width=2pt, <->, >=stealth]
        (checker) edge node [above] {Approve/Discuss/Reject} (output)
        (backend) edge node [above] {} (libraries)
        (risk) edge node[xshift=-4mm, yshift=-4mm] {Reads} (backend);
        \draw (researcher.-35) [bend right=20] edge[black, dashed, line width=2pt, ->, >=stealth] node[yshift=-5mm] {Finalise} (frontend.240);
        \draw (backend.230) edge [black, dashed, line width=2pt, ->, >=stealth] (output.145);
        \draw (frontend.-20) edge[black, dashed, line width=2pt, ->, >=stealth] node {} (backend.198);
    \end{tikzpicture}}
    \caption{Schematic illustration of SACRO. Disclosure checks are automated. Mitigation strategies optionally applied by researchers and approved by TRE staff.}
    \label{fig:schematic}
\end{figure*}

A significant advantage of the semi-automated principles-based approach is that researchers maintain control over the SDC process and can select the most appropriate mitigation strategies with approval from checkers, as opposed to the strict and brittle automatic rules-based approaches that are currently in use~\cite{Ritchie:2015}.

The operational design philosophy is extensively documented by Green~et~al.~\cite{Green:2020} who studied the characteristics that a semi-automated SDC solution needs to have to be feasible, effective, and a positive choice for users. The essential criteria are that it should be:

\begin{enumerate}
    \item Free and open source: the SACRO tools are available on GitHub under MIT License at: \url{https://github.com/AI-SDC/ACRO}.

    \item Easy to obtain: in addition to the source code available on GitHub, a Python ACRO package is available through the Python Package Index (PyPI), which provides a simple management system for distributing Python libraries and facilitating continuous updates. Similarly, the R ACRO package~\cite{acro-cran} is available through the Comprehensive R Archive Network (CRAN). ACRO and the SACRO viewer GUI are cross-platform, supporting GNU/Linux, macOS, and Windows; with v0.4.7 supporting Python 3.9--3.13.

    \item Easy to use: SACRO reduces barriers to adoption via a front-end application programming interface (API) that is identical to those already commonly used by researchers in their favoured language. Researchers prepare their data and statistical queries in the usual way, in their preferred language. When they are satisfied with the format of a query that generates, for example, a table or a regression, they then repeat it using the same commands prefixed by \code{acro}. The lightweight translation functions then call the Python back-end, which executes the queries and performs the requisite output checks. Easy to understand help and documentation is available within the GitHub repository.

    \item Comprehensive: ACRO identifies potentially disclosive outputs with a range of commonly used disclosure tests, applies mitigation strategies as requested, and flags outputs that have failed any tests. The current version of ACRO implements suppression and rounding as options for researchers to protect privacy, which are the most common forms of disclosure mitigation strategy.

    \item Transparent: identifying and optionally applying disclosure mitigation where requested, but with clear and rapid explanations that enable (i) researchers to understand and revise their requests; and (ii) an auditable record for TRE staff with simple summary reports to streamline their workflow; ACRO provides researchers with immediate feedback by displaying the results of both the checks and the output to the researcher, with the option to apply mitigations where appropriate. Details of the queries and results are stored in a list, which may subsequently be written to file for review by a human output checker. ACRO gives researchers control over the outputs that are submitted for review, e.g., the removal of unwanted outputs and choice of output format; currently JSON~\cite{JSON} or Microsoft Excel\textsuperscript{\tiny\textregistered}.

    \item Able to support exceptions under principles-based regimes; ACRO enables researchers to add comments to requested outputs and ensures that researchers supply supporting evidence for all outputs that fail any of the tests.

    \item Consistent: providing the same results across different studies within a TRE, and across TREs; ACRO contains a single back-end code base constituting a single source of truth for performing checks with extensibility for different languages and ongoing support and consistency.

    \item Scalable over users and outputs. Since ACRO is lightweight, session-based, and runs on the machine that generates outputs, it scales easily to many users and outputs.

    \item Able to implement an organisation's business rules for primary and secondary disclosure, which may vary across datasets or users. Within ACRO, SDC heuristics and parameters can be uniquely configured through a simple YAML~\cite{YAML} file specifying the risk appetite.

    The current version of ACRO does not yet address secondary disclosure (such as checking for differencing across tables), for two reasons. First, business rules for secondary checking are often not clear or comprehensive. Second, ACRO works by intercepting commands and assessing disclosure risk at the time the output is being produced. Analysing results post-hoc is a considerably harder problem, requiring the researcher to produce a lot more information and also locate the other outputs to be compared. 
\end{enumerate}
 
The SACRO toolkit is a product of the ``Semi-Automated Checking of Research Outputs'' project funded by Data and Analytics Research Environments UK (DARE UK).\footnote{\url{https://dareuk.org.uk/driver-project-sacro/}} This work builds upon a previous Eurostat-funded project, in which Green~et~al.~\cite{Green:2021} developed a proof-of-concept prototype for the proprietary Stata software. As part of the SACRO project, the SACRO team committed to review and re-develop the theory and operational guidelines for output SDC\@. The aim was threefold; first, to bring together key points from the SDC literature (and fill in some of the theoretical gaps) to provide an integrated guide to both theory and practice of output checking; second, to develop a new approach to output SDC based on classifications into groups; for details, see Derrick~et~al.~\cite{Derrick:2023}; third, to explicitly link theory to operational rules and their implementation in manual and automatic checking regimes.

Five UK TREs (OpenSAFELY\footnote{\url{https://www.opensafely.org}} at the University of Oxford and four Scottish Safe Havens) were involved in the SACRO project to provide detailed feedback on user and output checker perspectives; OpenSAFELY also took the lead in the design of the GUI\@. This group also directly tested the feasibility of installing and allowing the Python code to run on their systems as TREs differ in their perceptions of Python's `riskiness'. Moreover, the SACRO team contacted a large number of TREs in the UK and abroad, and set up a network of interested parties willing to be testers. Several engagement events with this group identified how they worked and what they would expect from a semi-automatic solution.

Subsequently, a `Community of Interest' group has been established to provide on-going peer support and new feature prioritisation. Currently, the mailing list\footnote{\url{mailto:sdc-reboot@jiscmail.ac.uk}} has over 50 subscribers from more than 25 different organisations. To illustrate the breadth of uptake, just some of those who are deploying SACRO and encouraging its use by researchers include: eDRIS/Public Health Scotland and the Scottish National Safe Haven; Clinical Practice Research DataLink (Medicines \& Healthcare Products Regulatory Agency, England), and the Secure Anonymised Information Linkage (SAIL) Databank (Wales).
 
The remainder of this article is organised as follows. Section~\ref{section:background} discusses the current state-of-the-art for output SDC with the Five Safes framework and SDC implementations. Section~\ref{section:acro} describes the SACRO toolkit, including the SDC checks implemented, as well as how to install and use the packages. Section~\ref{section:gui} provides details of the GUI for viewing and managing the results of SDC produced by ACRO\@. Finally, Section~\ref{section:future} discusses future plans.
 
\section{BACKGROUND}\label{section:background}

\subsection{FIVE SAFES AND OUTPUT DISCLOSURE CONTROL}

The Five Safes data governance framework~\cite{Ritchie:2017} is a set of principles that enable services to provide safe research access to their data and has been adopted by a range of TREs, including the UK Office for National Statistics (ONS), Health Data Research UK (HDR UK), and the National Institute for Health Research Design Service (NIHR), as well as many others worldwide.

The Five Safes are composed of:
\begin{itemize}
    \item {\bf Safe data}: (pseudo)anonymisation and minimisation of data provided to researchers.
    \item {\bf Safe projects}: research projects have been approved and make appropriate use of data.
    \item {\bf Safe people}: researchers trained and authorised to use data safely.
    \item {\bf Safe settings}: a secure environment that prevents un\-authorised use and ingress or egress of data other than through approved channels.
    \item {\bf Safe outputs}: screened and approved outputs that are non-disclosive.
\end{itemize}

TRE staff commonly follow the SDC handbook by Griffiths~et~al.~\cite{Griffiths:2019}, which provides advice on how to assess specific statistical outputs within the context of the Five Safes, as well as advice to organisations about how to set up SDC systems. Additionally, Jefferson~et~al.~\cite{Jefferson:2022} provide recommendations for the SDC of machine learning models trained within TREs. However, ensuring `safe outputs' is a complex and often costly human labour-intensive process. As DARE UK~\cite{DARE:2022} have recently highlighted: \emph{``there are already significant issues with staffing resources to support statistical disclosure control (safe outputs). This staffing issue is acting as a barrier to scaling up the use of TREs.''}

\subsection{SAFE VS.\ UNSAFE STATISTICS}

Ritchie~\cite{Ritchie:2008} first introduced the concept of `safe statistics', which aims to classify statistics as `safe' or `unsafe'. `Unsafe' statistics, such as tables of frequencies, present a number of potential disclosure risks. `Safe' statistics in contrast, such as the coefficients of correlation from model estimates, have no effective risk.

Five mitigation strategies are frequently employed to protect `unsafe' outputs: (i) generalisation, which reduces the specificity of a sensitive value; (ii) suppression, which hides sensitive values completely; (iii) permutation, where records are partitioned and the values of sensitive attributes are shuffled within subgroups; (iv) perturbation, where sensitive values are replaced with synthetically generated data; and (v) anatomisation, where sensitive values are split into separate tables to break any associations~\cite{Majeed:2020}.

Commonly, a minimum threshold rule is applied to the number of observations used by a statistic to ensure that there is sufficient uncertainty with respect to any individual respondent~\cite{Alves:2020}.

Dominance rules protect large respondent values from being approximated where the contribution to a statistic is dominated by only a few individuals. For example, the $p$\%-rule sorts the $N$ observations by magnitude and checks whether the sum of the smallest $N-3$ observations is at least $p$\% of the largest observation. The $NK$ rule checks that the largest $N$ observations contribute less than $K$\% of the total. Not all aggregation statistics are considered safe; typically, reporting minima or maxima values of a subgroup are prohibited.

Regressions are generally protected by checking that the residual degrees-of-freedom exceeds a minimum threshold. The purpose is to prevent an equation from masquerading as a model. Residual degrees of freedom is defined as the number of observations minus the number of restrictions embodied in the test or model. For example, for a simple linear regression it is $N-K$, where $K$ is the number of coefficients including the intercept; for chi-square it is $N$ test degrees of freedom. Following Brandt~et~al.~\cite{Brandt:2009}, 10 is typically used as the minimum. Categorical data is also frequently protected by enforcing minimum counts of the sample sizes, and applying rounding rules based on the significant digits.

Devising rules for each individual statistic used by analysts is infeasible, however it is possible to combine statistics into groups based not on statistical relation, but on common disclosure risks and solutions. Consequently, Ritchie~et~al.~\cite{Ritchie:2023} propose the use of `statistical barns' (statbarns) as a means to classify statistics for disclosure control purposes.

A statbarn is a collection of statistics that share the same characteristics for disclosure control purposes. That is, their mathematical form is similar; they share the same risks; they share the same responses to those risks; and output checking rules are applicable to all. For example, pie charts, histograms, and scatter plots are all forms of frequency table. Table~\ref{table:statbarns} lists the 14 statbarns that have been identified, with 12 classified for output checkers.

\begin{table}[t]
\caption{Statistical Barns Proposed by Ritchie et al.~\cite{Ritchie:2023}}
\label{table:statbarns}
    \centering
    \begin{tabular}{l l l}
        \toprule
        Barn & Example & Class\\
        \midrule
        Frequencies & Frequency tables & Unsafe\\
        Statistical hypothesis tests & $t$-stats, $p$-stats, $f$-stats & Safe\\
        Correlation coefficients & Regression coefficients & Safe\\
        Position & Median, quartiles & Unsafe\\
        End points & Maximum, minimum & Unsafe\\
        Shape & s.d., skewness, kurtosis & Safe\\
        Linear aggregations & Means, totals & Unsafe\\
        Mode & Mode & Safe\\
        Nonlinear concentration ratios & Herfindahl index & Safe\\
        Calculated ratios & Odds and risk ratios & Unsafe\\
        Survival tables & Hazard/survival tables & Unsafe\\
        Gini coefficient & Gini coefficient & Safe\\
        Linked/multi-level tables & Nested categorical data & ?\\
        Clusters & Cluster analysis & ?\\
        \bottomrule
    \end{tabular}
\end{table}

\subsection{PRINCIPLES-BASED DISCLOSURE CONTROL}

Traditionally, TREs have focused on rules-based approaches to output SDC, where a rule is a hard limit with no exceptions permitted. More recently, it has been recognised that a principles-based approach to SDC is more effective since defining hard rules that are neither too restrictive nor too loose is extremely difficult~\cite{Ritchie:2015}. For example, a frequency table with zeros in some cells is highly disclosive if they reveal that no one in a given region earned more (or less) than $k$~USD\@. In contrast, reporting that no one born without organ $k$ was diagnosed with cancer of said organ would usually be recognised as a ``structural zero'', and permitted.

In principles-based SDC, both researchers and output checkers undertake SDC training. They are provided with a set of heuristics rather than strict, definitive rules. These heuristics serve as initial guidelines. If a researcher wishes to request approval for outputs that deviate from these guidelines, they must meet three conditions: (i) ensure the outputs do not disclose sensitive information; (ii) demonstrate the outputs' significance; and (iii) justify why the request is exceptional. The responsibility lies with the researcher to demonstrate that any potentially risky outputs do not disclose sensitive information; although the final decision rests with the checker. Since there are no strict rules, both the researcher and the checker need expertise in understanding disclosure risks and exercising judgement.

Semi-automated output checking therefore aims to improve the rigour and consistency of the output disclosure control process and reduce human workload by automatically applying best-practice principles-based SDC techniques. That is, automatically identifying, reporting, and optionally applying mitigation strategies to disclosive outputs where possible.

The use of semi-automated SDC tools for checking research outputs therefore enables:
\begin{itemize}
    \item An explicit change in the dynamics so that SDC is something done with researchers rather than to them.
    \item Reduced costs for TREs.
    \item Improved efficiency with skilled personnel focusing on the areas of most significant risk. 
    \item Improved security and public confidence by reducing the chance of human error.
    \item Improved auditing by generating structured outputs.
    \item Improved user experience by making output checking quicker.
    \item Increased consistency within and between TREs.
\end{itemize}
 
\subsection{TOOLS FOR SDC}

A small number of SDC tools have been produced to assist in the process of achieving `safe outputs', such as sdcTable~\cite{sdcTable} and tauArgus~\cite{tauArgus} by Statistics Netherlands; however they require expert knowledge of SDC to use effectively. Moreover, they are exclusively designed for tabular (frequency and magnitude) outputs, and do not cover the range of statistics produced by researchers. They are predominantly used for regularly repeating reports in National Statistics Institutes. Metadata must be created to describe each output and control parameters configured. The need to rewrite the metadata for each table makes these tools poorly suited for research use. However, for scenarios where the same tables are repeatedly generated and secondary differencing is considered a significant problem, the investment in setting up the tools can be cost-effective.

Additionally, Statistics Netherlands have developed sdcMicro~\cite{sdcMicro}, which is an R package and GUI that provides various methods for anonymising data and performing risk estimation.

DataSHIELD\footnote{\url{https://datashield.org}} is an infrastructure and suite of R packages~\cite{datashield} that aims to enable remote and non-disclosive analysis of distributed sensitive research data while avoiding the normal practice of human involvement in output checking. It operates under the principle that researchers never see or have access to the underlying data, even for manipulation. This is achieved by providing a set of restricted bespoke commands for performing data manipulation and querying. This approach implements rules-based rather than principles-based SDC and some rules (e.g., allowing min/max values, and not suppressing table cells with zero counts) differ significantly from the standard practice as described in the SDC handbook~\cite{Griffiths:2019}.

The Research Data and Service Centre (RDSC) of the Deutsche Bundesbank provide two tools to allow researchers to check whether the results generated in their research project comply with the RDSC's SDC rules~\cite{Blaschke:2022}. Researchers run a special command directly after generating their results and get immediate feedback. The RDSC provide nobsdes5/nobsreg5, which are a set of Stata ado files, and the sdcLog~\cite{sdcLog} package, which provides a set of new R functions. These tools help researchers in checking descriptive statistics and models, and in calculating extreme values that are not individual data. However, since these tools implement new commands, researchers must learn them and adapt their workflow.

With the aim of developing a more general semi-automated solution for use by researchers in TREs, improving the efficiency of the process, and (where applicable) reducing the amount of user training required, a recent Eurostat project~\cite{Green:2021} developed a proof-of-concept prototype in Stata where primary disclosure is regulated by a set of simple rules. Importantly, the design philosophy was of like-for-like replacement using the same call signatures as common Stata commands, but prefixed by a special keyword. Here we build upon the lessons learned from developing that prototype with the aim of expanding its coverage and impact whilst retaining a minimal learning curve. We achieve this by adopting the principles-based approach to SDC outlined by Ritchie~et~al.~\cite{Ritchie:2023} and develop the SACRO toolkit, which is primarily implemented in Python, but with cross-language support for R and Stata, using analytical interfaces familiar to researchers.

\section{THE SACRO TOOLKIT}\label{section:acro}

\subsection{OVERVIEW}

The SACRO toolkit consists of (i) the ACRO Python package, which implements a range of automated disclosure tests and optional mitigation algorithms; (ii) additional language packages that create their own Python virtual environment and provide wrapper functions that interface with the ACRO Python package; and (iii) a GUI that uses the reports generated to assist staff checkers in performing their reviews. Since the additional language packages use the same Python back-end for running and reporting the tests, the results are consistent across different platforms and languages. In addition, the same configuration files can be used across different languages and the reports generated for use with the GUI are identical.

As the additional language packages depend on the core Python back-end, they are pinned to a specific release of the Python ACRO package. This enables updates to be made to the Python package, including potentially breaking API changes, without affecting dependant languages. The additional language packages are then updated to incorporate these changes.

Since the SACRO toolkit has been designed as a drop-in replacement in order to maintain familiar interfaces, it necessarily depends on particular versions of the analysis tools supported. These versions are similarly pinned to specific ranges of versions to maintain stability and interoperability, and kept under continuous review. The versions currently supported are noted within the documentation.

A CHANGELOG markdown file is included in the Python GitHub repository which maintains summaries of all updates made for each version of the tool; and a NEWS markdown file similarly records changes made for the R package. 

\subsection{DISCLOSURE CONTROL IMPLEMENTED}\label{section:sdc}

For tabular data (e.g., cross tabulation and pivot tables), the ACRO tool prohibits the reporting of the maximum or minimum value in any cell that represents a sub-group of one or more contributors. Moreover, ACRO reports the reason why the value of an aggregation statistic (mean, median, variance, etc.) for any cell is deemed to be \textit{sensitive}.

The current version of ACRO supports the three most common tests for sensitivity: ensuring the number of contributors is above a frequency threshold, and testing for dominance via $p\%$ and $NK$ rules. Sensitive cells may optionally be suppressed. Currently, primary suppression with adjustment of marginal totals is performed. This is flagged for the output checkers' attention to assess the potential risk of differencing. In addition, any outputs found to include negative or missing values are automatically flagged for human review since the results of SDC are not well defined in these circumstances.

For tables, ACRO builds a series of masks that indicate which cells fail the sensitivity tests for each check. For reach check there are two stages; first a version of the requested table is built using a custom aggregation function (or `count' for cell thresholds). The resulting tables are then binarised via comparison with a parameter that represents the data-owners' \textit{appetite} for that risk. A summary outcome table indicating which rule was applied to each cell is presented to the researcher, along with the result of the query. If suppression is enabled, the offending cells are censored.

For regressions such as linear, probit, and logit, the tests verify that the number of residual degrees of freedom exceeds a threshold. There is currently limited support for graphical plots, however ACRO supports the SDC of histograms using the same rules as for frequencies since these fall under the same statbarn category; if a histogram is found to be disclosive and suppression is enabled, the plot is hidden from the researcher. Additionally, survival plots such as Kaplan-Meier may be protected using a rounding mitigation strategy.

As noted above, all of these tests and checks are configurable according to the TRE's risk appetite. The data custodian, e.g., TRE staff member, specifies the parameter values used for the output checks in a YAML configuration file, which is loaded upon ACRO initialisation. The default ACRO parameters are shown in Table~\ref{table:params}. Future releases will offer the option to configure parameters on a dataset or attribute level, in addition to the current session-based method.

\begin{table*}[t]
\caption{Default Parameters for Sensitivity Tests}
\label{table:params}
    \centering
    \begin{tabular}{l l r}
        \toprule
        Description & Parameter & Value \\
        \midrule
        Minimum frequency threshold for tabular data & \code{safe\_threshold} & 10.0 \\
        Minimum degrees-of-freedom for analytical statistics & \code{safe\_dof\_threshold} & 10.0 \\
        $N$ parameter in $NK$ test & \code{safe\_nk\_n} & 2.0 \\
        $K$ parameter in $NK$ test & \code{safe\_nk\_k} & 0.9 \\
        Minimum ratio for $p$\% test & \code{safe\_pratio\_p} & 0.1 \\
        Whether to check for missing values & \code{check\_missing\_values} & False \\
        Frequency thresholds for survival tables and plots & \code{survival\_safe\_threshold} & 10.0 \\
        Whether to consider zeros to be disclosive & \code{zeros\_are\_disclosive} & True \\
        \bottomrule
    \end{tabular}
\end{table*}

\subsection{PYTHON PACKAGE}\label{section:python}

Python is a popular multi-platform language widely used for data analysis and machine learning. PyPI provides a simple and easy to use package management system for distributing open source Python libraries. Pandas~\cite{pandas} and the statsmodels~\cite{statsmodels} Python libraries are mature, popular, and well-supported packages for data analysis, statistical testing, and statistical data exploration. Pandas is used by more than 55\% of all Python users~\cite{JetBrains:2021}.

The use of Python as the primary implementation therefore enables the leveraging of existing expertise and community support with these packages so that the ACRO front-end can be as similar to the API researchers already know and trust, and further facilitates the rapid development of disclosure checking functionality on the back-end.

As the PyPI distribution system is simple and allows the use of semantic versioning, it supports a rapid and iterative develop-and-deploy strategy to provide continuing functionality and improvements. The Python ACRO package also depends on Matplotlib~\cite{Hunter:2007} to support data visualisations, NumPy~\cite{numpy} as a core numerical matrices package, and PyYAML~\cite{PyYAML} for YAML configuration file support.

After installing ACRO from PyPI, a \textit{session} can be started by instantiating an \class{ACRO} object as seen in Figure~\ref{fig:pyinit}, which is used to store and maintain outputs. Each session would typically be contained within a single period of (physical or virtual) access to a TRE\@. The \class{ACRO} constructor optionally accepts a string parameter \code{config} that can be used to specify an alternative YAML rule configuration file, and a Boolean parameter \code{suppress} that specifies whether suppression should be applied within the session.

\begin{figure}[t]

\begin{terminal}
>>> from acro import ACRO
>>> acro = ACRO(suppress=True)
\end{terminal}

\info{
    \begin{tabular}{l}
        INFO:acro:version: 0.4.7\\
        INFO:acro:config: \{`safe\_threshold': 10,\\ `safe\_dof\_threshold': 10,
        `safe\_nk\_n': 2, `safe\_nk\_k': 0.9,\\ `safe\_pratio\_p': 0.1,
        `check\_missing\_values': False,\\ `survival\_safe\_threshold': 10,\\
        `zeros\_are\_disclosive': True\}\\
        INFO:acro:automatic suppression: True\\
    \end{tabular}
}

\caption{Example initialisation of ACRO in Python.}%
\label{fig:pyinit}

\end{figure}

Standard data analysis operators can then be used by pre-fixing the function with \class{acro}. Figure~\ref{fig:crosstab} shows how the \class{pandas.crosstab} function can be used with the same arguments and in the same way, operating on a \class{pandas.DataFrame}, \code{df}. However, in this instance, the \code{R/G} column and a single cell in column \code{N} is automatically suppressed since it violates disclosure checks and mitigation has been enabled. The output also includes an explanation of which disclosure checks failed for each cell, a summary of the checks, and the name assigned to the output so that it can be referenced subsequently; for example to add comments or request an exception from a human checker. 

\begin{figure}[t]

\begin{terminal}
>>> table = acro.crosstab(df.year, df.grant_type,
...  values=df.inc_grants, aggfunc="mean")
\end{terminal}

\info{
    \begin{tabular}{l r r r r}
        \multicolumn{5}{l}{INFO:acro:get\_summary(): fail;}\\

        \multicolumn{5}{l}{threshold: 6 cells suppressed; p-ratio: 2 cells suppressed;} \\
        \multicolumn{5}{l}{nk-rule: 1 cells suppressed;} \\
        \multicolumn{5}{l}{INFO:acro:outcome\_df:} \\
        grant\_type & G & N & R & R/G \\
        \multicolumn{5}{l}{year} \\
        2010 & ok & p-ratio; & ok & threshold; p-ratio; nk-rule; \\
        2011 & ok & ok & ok & threshold; \\
        2012 & ok & ok & ok & threshold; \\
        2013 & ok & ok & ok & threshold; \\
        2014 & ok & ok & ok & threshold; \\
        2015 & ok & ok & ok & threshold; \\
        \multicolumn{5}{l}{INFO:acro:records:add(): output\_0} \\
    \end{tabular}
}\\

\code{
    \begin{tabular}{l r r r r}
        grant\_type & G & N & R & R/G \\
        \multicolumn{5}{l}{year} \\
        2010 & 9921906.0 & NaN & 8402284.0 & NaN \\
        2011 & 8502246.0 & 124013.859375 & 7716880.0 & NaN \\
        2012 & 11458580.0 & 131859.062500 & 6958050.5 & NaN \\
        2013 & 13557147.0 & 147937.796875 & 7202273.5 & NaN \\
        2014 & 13748147.0 & 133198.250000 & 8277525.0 & NaN \\
        2015 & 11133433.0 & 146572.187500 & 10812888.0 & NaN \\
    \end{tabular}
}
    \caption{Example ACRO crosstab. Disclosive cells suppressed with NaN.}%
    \label{fig:crosstab}
\end{figure}

Similarly, pivot tables can be created as in Figure~\ref{fig:pivot}. Note that the only API difference is calling \class{acro.pivot\_table} instead of \class{pandas.pivot\_table}. In this example, the entire column \code{R/G} has been withheld as a result of enabling suppression. Since the tables in these examples failed SDC, exception requests must be provided, as in Figure~\ref{fig:exception}.

\begin{figure}[t]

\begin{terminal}
>>> table = acro.pivot_table(
...   df, index=["year"], columns=["grant_type"],
...   values=["inc_grants"], margins=True, aggfunc="sum")
\end{terminal}

\info{
    \begin{tabular}{l r r r r}
        \multicolumn{5}{l}{INFO:acro:get\_summary(): fail;}\\
        \multicolumn{5}{l}{threshold: 6 cells suppressed; p-ratio: 2 cells suppressed;}\\
        \multicolumn{5}{l}{nk-rule: 1 cells suppressed;}\\

        \multicolumn{5}{l}{INFO:acro:outcome\_df:}\\
        & \multicolumn{4}{l}{inc\_grants} \\
        grant\_type & N & R & R/G & All \\
        year & & & & \\
        2010 & p-ratio & ok & threshold; p-ratio; nk-rule; & ok \\
        2011 & ok & ok & threshold; & ok \\
        2012 & ok & ok & threshold; & ok \\
        2013 & ok & ok & threshold; & ok \\
        2014 & ok & ok & threshold; & ok \\
        2015 & ok & ok & threshold; & ok \\
        All  & ok & ok & ok & ok \\
        \multicolumn{5}{l}{INFO:acro:Disclosive cells were deleted from the dataframe} \\
        \multicolumn{5}{l}{before calculating the pivot table} \\
        \multicolumn{5}{l}{INFO:acro:records:add(): output\_1} \\
    \end{tabular} 
    }\\

\code{
    \begin{tabular}{l r r r}
        & inc\_grants & & \\
        grant\_type & N & R & All \\
        year & & & \\                                                          
        2010 & NaN & 5.041371e+08 & 6.430438e+08 \\
        2011 & 7192804.0 & 5.324647e+08 & 6.671912e+08 \\
        2012 & 7779685.0 & 4.801055e+08 & 6.597639e+08 \\
        2013 & 8728330.0 & 5.113614e+08 & 7.234470e+08 \\
        2014 & 7858697.0 & 5.545942e+08 & 7.686751e+08 \\
        2015 & 8501187.0 & 5.514573e+08 & 6.935597e+08 \\
        All & 90060704.0 & 3.134120e+09 & 4.155681e+09 \\
    \end{tabular} 
}

    \caption{Example pivot table produced with ACRO. R/G column withheld.}
    \label{fig:pivot}
\end{figure}

\begin{figure}[t]

%

\begin{terminal}
>>> acro.add_exception(
...   "output_1", "Trust me, I'm a professor!")
\end{terminal}

\info{
    \begin{tabular}{l}
        INFO:acro:records:exception request was added to output\_1
    \end{tabular}
}

\caption{Example adding exceptions to ACRO outputs.}%
\label{fig:exception}

\end{figure}

A benefit of building ACRO on the vast user-base of Pandas is the plethora of help and guidance available. For example, should a researcher wish to create a table with survey-weighted data, a simple web query provides examples of how this can be done by defining a custom aggregation function, and passing that to the tabulation command. Note, however, that in this case the checks would be performed using the unweighted data.
 
In addition to common Pandas tabular operations, ACRO provides disclosure control tests for common statsmodels regression operations. Figure~\ref{fig:ols} shows an example where an ordinary least squares is performed on \(x\) and \(y\) where the residual degrees of freedom are checked. Note that the only API difference is calling \class{acro.ols} instead of \class{statsmodels.api.OLS}.

\begin{figure}[t]

\begin{terminal}
>>> results = acro.ols(y, x)
>>> results.summary()
\end{terminal}

\info{
    \begin{tabular}{l}
        INFO:acro:ols() outcome: pass; dof=807.0 >= 10 \\
        INFO:acro:records:add(): output\_2 \\
    \end{tabular}
}\\

\code{
    \fontsize{7pt}{7pt}\selectfont
OLS Regression Results \\

\setlength{\tabcolsep}{2pt}
\renewcommand{\arraystretch}{1}

\begin{tabular}{r r r r}
    Dep.\ Variable: & inc\_activity & R-squared: & 0.894 \\
    Model: & OLS & Adj.\ R-squared: & 0.894 \\
    Method:	& Least Squares & F-statistic: & 2276. \\
    Date: & Mon, 26 Aug 2024 & Prob (F-statistic): & 0.00 \\
    Time: & 23:05:54 & Log-Likelihood: & -14493. \\
    No.\ Observations: & 811 & AIC: & 2.899e+04 \\
    Df Residuals: & 807 & BIC: & 2.901e+04 \\
    Df Model: & 3 & & \\
    Covariance Type: & nonrobust & & \\
\end{tabular}\\ \\

\begin{tabular}{r r r r r r r}
    & coef & std err & t & P>|t| & [0.025 & 0.975] \\
    const & 3.994e+05 & 5.31e+05 & 0.752 & 0.452 & -6.43e+05 & 1.44e+06 \\
    inc\_grants & -0.8856 & 0.025 & -36.128 & 0.000 & -0.934 & -0.837 \\
    inc\_donations &  -0.6659 & 0.016 & -40.905 & 0.000 & -0.698 & -0.634 \\
    total\_costs & 0.8318 & 0.011 & 78.937 & 0.000 & 0.811 & 0.853 \\
\end{tabular}\\ \\

\begin{tabular}{r r r r}
    Omnibus: & 1348.637 & Durbin-Watson: & 1.424 \\
    Prob(Omnibus): & 0.000 & Jarque-Bera (JB): & 1298161.546 \\
    Skew: & 10.026 & Prob(JB): & 0.00 \\
    Kurtosis: & 197.973 & Cond.\ No.\ & 1.06e+08 \\
\end{tabular}

}
    \caption{Example ACRO ordinary least squares regression.}
    \label{fig:ols}

\end{figure}

Limited support for graphical plots is available. For example, Figure~\ref{fig:hist} shows the result of attempting to produce a disclosive histogram. In this case, suppression has been enabled and so the plot is therefore hidden from the researcher. Note that the only API difference is calling \class{acro.hist} instead of \class{pandas.hist}.

\begin{figure}[t]

\begin{terminal}
>>> hist = acro.hist(df, "inc_grants")
\end{terminal}

\info{
    \begin{tabular}{l}
        WARNING:acro:Histogram will not be shown as the inc\_grants\\ 
        column is disclosive. \\
        INFO:acro:status: fail \\
        INFO:acro:records:add(): output\_3 \\
    \end{tabular}
}

\caption{Example producing a disclosive histogram with ACRO.}%
\label{fig:hist}

\end{figure}

The \class{ACRO} class includes a number of session management functions to display, rename, remove, or add comments to existing outputs, so that the researcher maintains control over the outputs they wish to request. An example renaming an output for convenience is shown in Figure~\ref{fig:rename}; an example adding comments to existing outputs is shown in Figure~\ref{fig:comments}; and an example removing an unwanted output can be seen in Figure~\ref{fig:remove}.

\begin{figure}[t]

\begin{terminal}
>>> acro.rename_output("output_0", "my_crosstab")
\end{terminal}

\info{
    \begin{tabular}{l}
        INFO:acro:records:rename\_output():\\
        output\_0 renamed to my\_crosstab
    \end{tabular}
}

    \caption{Example renaming an ACRO output.}
    \label{fig:rename}
\end{figure}

\begin{figure}[t]

\begin{terminal}
>>> acro.add_comments("my_crosstab",
...  "This is a crosstab between year and grant_type")
\end{terminal}

\info{
    \begin{tabular}{l}
        INFO:acro:records:a comment was added to my\_crosstab
    \end{tabular}
}

\caption{Example adding comments to an ACRO output.}%
\label{fig:comments}

\end{figure}

\begin{figure}[t]

\begin{terminal}
>>> acro.remove_output("output_3")
\end{terminal}

\info{
    \begin{tabular}{l}
        INFO:acro:records:remove(): output\_3 removed \\
    \end{tabular}
}

    \caption{Example removing an unwanted ACRO output.}%
    \label{fig:remove}
\end{figure}

Moreover, as previously shown the researcher can request an exception for a particular output and include their supporting evidence; unsupported outputs can be added with the \code{add\_custom\_output} function and by specifying the path to any file. These unsupported outputs are then included in the final requested release and highlighted for the attention of a human checker. When the researcher is finished with their analysis, they simply call the \code{finalise} function.

If any of the outputs have failed the SDC checks at this stage and an exception has not already been requested, the researcher is prompted to provide one. An SDC report is then created in JSON and all outputs are written to a directory named ``outputs'' by default; a subdirectory with SHA-256 checksums~\cite{SHA} is also generated. Optionally, an alternative directory may be requested by supplying a string parameter \code{path} to \code{finalise}. Additionally, the report can be written in Excel\textsuperscript{\tiny\textregistered} format by passing a string parameter \code{ext} set to \code{xlsx}. Having \code{finalise} run automatically when the object is destroyed (i.e., session closes) was considered, but for now this was decided against since researchers may choose to save their work between sessions and not wish to go through the full release process.

The automatic suppression of outputs can be enabled via the Boolean \code{suppress} constructor parameter; by default it is disabled. Regardless of whether suppression is applied, the results of the checks are included in the final JSON report. This report file can then be ingested by the GUI described in Section~\ref{section:gui} to display the disclosure results and mitigation techniques can be applied on a case by case basis.

Within the ACRO Python package, the functionality of the \class{ACRO} class is split into a number of separate classes for maintainability and extensibility. A \class{Tables} class contains the code necessary to perform disclosure checks on tabular data, such as \code{crosstab}. A separate \class{Regression} class contains the code for checking regressions such as \code{logit} and \code{probit}. The \class{ACRO} class thus inherits from each of these classes so that the researcher can invoke all of the functions directly on the instantiated object, which tracks the outputs generated and writes them upon \code{finalise}.

As Internet access is restricted from within TRE (virtual) environments, the ACRO package has extensive docstrings to support Python's inbuilt help mechanisms; for example:

\begin{terminal}
>>> help(acro.ACRO)
\end{terminal}

Standard Python coding and naming practices have been used throughout~\cite{VanRossum:2001, Goodger:2001}. GitHub continuous integration (CI) runners automatically generate and publish API documentation using the Python docstrings written in numpydoc~\cite{numpydoc} format. The project also contains a configuration file for the pre-commit~\cite{pre-commit} tool, which is used to run mypy~\cite{mypy} type checking and enforce standards and code formatting with Ruff~\cite{Ruff}. A GitHub bot is configured to automatically run pre-commit when any pull requests or pushes are made to the main GitHub branch. Additionally, a CI runner is configured to perform static code analysis with pylint~\cite{pylint}.

Extensive pytest~\cite{pytest} unit tests have been written, which currently cover \(>99\)\% of the Python source code. A CI runner is configured to automatically execute these tests and report the coverage results for all pull requests and pushes made to the main branch.

Moreover, CI runners are configured to pip install the main branch and execute the tests with a range of Python versions on Ubuntu, macOS, and Windows operating systems for quality assurance whenever a pull request has been merged to the main branch. An additional CI runner is used for building and publishing ACRO to PyPI\@; and we have recently added the ability to generate artifact attestations~\cite{attest} to enable users to confirm the provenance of their ACRO packages.

Some Jupyter notebooks~\cite{Granger:2021} demonstrating example code usage and output are available within the ACRO project repository. The currently implemented methods (v0.4.7) are listed as follows; for more details see the ACRO API documentation built with the tool Sphinx~\cite{sphinx}, which is available: \url{https://ai-sdc.github.io/ACRO/}.

\subsubsection{PYTHON TABLE METHODS}

\begin{itemize}
    \item \code{crosstab}(index, columns[, values, rownames, \ldots])\\
        Compute a cross tabulation of two (or more) factors.\\
        API: \class{pandas.crosstab}.
    \item \code{hist}(data, column[, by\_val, grid, \ldots])\\
        Create a histogram from a single column.\\
        API: \class{pandas.DataFrame.hist}.
    \item \code{pivot\_table}(data[, values, index, columns, \ldots])\\
        Create a spreadsheet-style pivot table as a \class{DataFrame}.\\
        API: \class{pandas.pivot\_table}.
    \item \code{surv\_func}(time, status, output[, entry, \ldots])\\
        Estimate the survival function.\\
        API: \class{statsmodels.duration.survfunc.SurvfuncRight}.
    \item \code{survival\_plot}(survival\_table, survival\_func, \ldots)\\
        Create survival plot according to suppression status.
    \item \code{survival\_table}(survival\_table, safe\_table, \ldots)\\
        Create survival table according to suppression status.
\end{itemize}

\subsubsection{PYTHON REGRESSION METHODS}

\begin{itemize}
    \item \code{logit}(endog, exog[, missing, check\_rank])\\
        Fit logit model.\\
        API: \class{statsmodels.discrete.discrete\_model.Logit}.
    \item \code{logitr}(formula, data[, subset, drop\_cols])\\
        Fit logit model from a formula and \class{DataFrame}.\\
        API: \class{statsmodels.formula.api.logit}.
    \item \code{ols}(endog[, exog, missing, hasconst])\\
        Fit ordinary least squares regression.\\
        API: \class{statsmodels.regression.linear\_model.OLS}.
    \item \code{olsr}(formula, data[, subset, drop\_cols])\\
        Fit ordinary least squares regression from a formula and \class{DataFrame}.\\
        API: \class{statsmodels.formula.api.ols}.
    \item \code{probit}(endog, exog[, missing, check\_rank])\\
        Fit probit model.\\
        API: \class{statsmodels.discrete.discrete\_model.Probit}.
    \item \code{probitr}(formula, data[, subset, drop\_cols])\\
        Fit probit model from a formula and \class{DataFrame}.\\
        API: \class{statsmodels.formula.api.probit}.
\end{itemize}

\subsubsection{PYTHON SESSION MANAGEMENT METHODS}

\begin{itemize}
    \item \code{ACRO}([config, suppress])\\
        Create an \class{ACRO} object for session management.
    \item \code{add\_comments}(output, comment)\\
        Add a comment to an output.
    \item \code{add\_exception}(output, reason)\\
        Add an exception request to an output.        
    \item \code{custom\_output}(filename[, comment])\\
        Add an unsupported output to the results list.
    \item \code{finalise}([path, ext])\\
        Create a results file for checking.
    \item \code{print\_outputs}()\\
        Print the current results list.
    \item \code{remove\_output}(key)\\
        Remove an output from the results list.
    \item \code{rename\_output}(old, new)\\
        Rename an output.
\end{itemize}

\subsection{R PACKAGE}\label{section:r}

The R ACRO package~\cite{acro-cran} is an example of cross-language support and is available through CRAN\@. It provides a set of R wrapper functions that execute Python back-end checking within a virtual environment via the reticulate package~\cite{reticulate}, which provides automatic conversions for types such as the R \class{data.frame} to Pandas \class{DataFrame}. The code is available under MIT license at: \url{https://github.com/AI-SDC/ACRO-R}.

GitHub CI runners automatically validate the package for CRAN by running \code{R CMD check}~\cite{R-CMD} using different operating systems and Python versions to ensure cross-platform operability. CI runners also generate and publish API documentation using pkgdown~\cite{pkgdown} and generate test code coverage reports.

For regressions, the common R \code{lm} and \code{glm} functions were shadowed with equivalent versions implemented as \code{acro\_lm} and \code{acro\_glm}, respectively, exploiting the fact that statsmodels already supports R-style specification of formulae. For tabular data, the standard R \code{table} command has been shadowed with \code{acro\_table} to perform cross tabulation analysis.

While the dplyr package~\cite{dplyr} is commonly used, no simple pivot table functions are provided; instead various combinations of \code{groupby} and \code{summarize} etc.\ are used. Therefore, at this stage of development, the Python cross tabulation and pivot table functions were directly interfaced with \code{acro\_crosstab} and \code{acro\_pivot\_table}.

To start a session, instantiate an \class{ACRO} object:

\begin{terminal}
>>> library("acro")
>>> acro_init(suppress = TRUE)
\end{terminal}

Subsequently, the Python package functions may be used, however for simplicity the package provides R helper functions that use underscore naming. For example, the \code{crosstab} function can be used identically to the Python version by calling \code{acro\_crosstab}. Similarly, the \code{pivot\_table} function becomes \code{acro\_pivot\_table}, etc. See the ACRO R reference manual on CRAN for further details.\footnote{\url{https://cran.r-project.org/web/packages/acro/acro.pdf}}
 
\subsection{STATA INTERFACE}\label{section:stata}

The ACRO Stata interface is implemented for versions below 15 and for versions 17 and above when the syntax changed. It makes extensive use of Stata's SFIToolkit~\cite{SFIToolkit} to manage a Python session, transfer data in memory from Stata to a Pandas \class{DataFrame} in the Python session, and results back to the Stata window.

A simple \class{acro.ado} file defines a new function \code{acro} which takes as parameters either one of the \class{ACRO} session management methods (adding \code{init} to start a session) or the name of a standard Stata function such as \code{table}, \code{regress}, etc. Stata's inbuilt parsing functions are used to separate out the parts of command and pass them as lists to a Python function \code{parse\_and\_run} which handles the rest of the translation between the two languages.

The Stata ado files implementing the ACRO interface can be found in the `stata' folder located within the main ACRO GitHub repository.\footnote{\url{https://github.com/AI-SDC/ACRO/tree/main/stata}} The files should be placed in the standard Stata ado file directory and the Python ACRO package must be installed, which includes a Stata to Python parser.

\section{GRAPHICAL USER INTERFACE}\label{section:gui}

OpenSAFELY developed a platform-independent stand-alone tool for checkers to view the outputs produced by ACRO, understand the risks associated with each output, make audited decisions, and produce zipped packages of files for release.

The viewer consists of (i) a Django~\cite{django} web application with a vanilla JavaScript user interface that renders a set of ACRO outputs for review; and (ii) an Electron~\cite{electron} application and installer that bundles the web application. It is available on GitHub under GNU General Public License v3 at \url{https://github.com/AI-SDC/SACRO-Viewer}.

The application is packaged using standard tooling, e.g., as an msi for Windows, deb for Debian/Ubuntu, and dmg for macOS operating systems; all dependencies are bundled with the installers, including a chrome-based web browser, Python, and JavaScript runtimes. The Cypress~\cite{cypress} Javascript framework is used to implement a number of automated user interface tests for CI\@. The viewer supports and renders a range of different file types for results from queries not yet supported by ACRO\@. Hence the viewer can be used for making and recording decisions, even if the researcher has not used ACRO during their analysis; automated disclosure risk analysis is not provided in those cases.

To view outputs, the user opens a directory containing the ACRO output files. The viewer then detects if there is ACRO generated metadata and uses that to display the files. If no ACRO data is available, it automatically generates it, adding each file in the directory as a custom output. The GUI viewer includes the ability to view relevant source code snippets and syntax highlighting to aid human review.

\begin{figure}[t]
    \includegraphics[width=1\columnwidth]{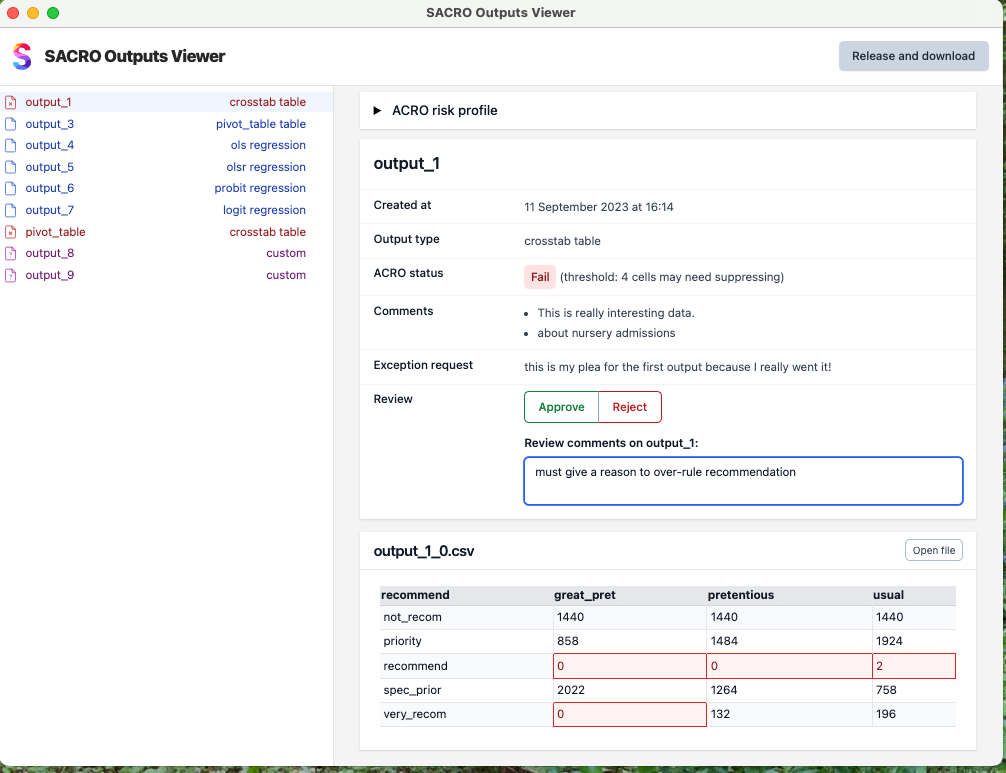}
    \caption{Screenshot of the SACRO viewer.}%
    \label{fig:gui}
\end{figure}

A screenshot of the GUI can be seen in Figure~\ref{fig:gui}. The left pane displays the list of outputs to review. The top right pane provides an option to view the ACRO risk profile. The centre right pane shows details of the currently selected output, including comments from the researcher, any exception requests, and the ACRO recommendation.

Moreover, the GUI provides buttons for checkers to approve or reject the output and supply feedback, e.g., if the checker wishes to override the ACRO recommendation. The bottom right pane displays the currently selected output with any specific components that violate SDC checks highlighted in red. Hovering over the tables displays pop-up information detailing which test failed for each cell. When checking has been completed, the top right `Release and download' button can be used to record the overall comments and create a (zip) package for release, which automatically excludes any rejected outputs.

\section{FUTURE PLANS}\label{section:future}

Future plans include the ongoing development of both the main Python package and additional translation languages to support more analyses and enhance functionality. Improvements will also result from extending the coverage of the R Tidyverse~\cite{tidyverse} family such as dplyer, and the production of an easy to install package for Stata.

The user experience will be enhanced by facilitating the selective application of disclosure mitigations on a query by query basis to provide even more control over the process; implementing additional mitigation options such as rounding and differential privacy for tables and statistical models; improved support for graphical plots; and the ability to configure rules at a dataset or even attribute level.

Currently, the results of SDC checks for tables are presented by displaying the whole table, where the reasons for failure are presented within each cell; while this works well for small to medium size tables, it can be overwhelming for larger tables. Presenting the row/column numbers of the offending cells with an explanation of the SDC check instead of the entire table will aid comprehension.

Additional parameterised disclosure checks will increase the effectiveness of the toolkit, such as testing for minimum sample sizes in the use of regressions applied to categorical data. As previously mentioned, the current version of the SACRO toolkit does not implement secondary disclosure checks. However, the standardised format of outputs and summary documents enables the manual SDC review process to identify potential differencing attack risks across multiple releases or researchers' analyses. Future development aims to implement a standardised mechanism to systematically leverage these summary documents over time, flagging potential differencing risks across research sessions and creating a searchable library of outputs to enhance assessment of secondary disclosure risks, even if initially partial.

Moreover, future versions of the GUI viewer may include outputs from checking trained machine learning models, such as those from the sacroml~\cite{sacroml} Python package. The GUI viewer has been designed for checkers to review the results produced by researchers; the development of a GUI for researchers will enable them to view, modify, and supply information to checkers. Furthermore, as the TRE infrastructure landscape evolves, and new lightweight dashboard frameworks such as Streamlit~\cite{streamlit} increase adoption, the question of choice of tools for implementing the viewer design inevitably remain open. 
 
Future work also includes the development of a toolkit that can be embedded within projects such as OpenSAFELY and medical TREs. Additional features and improved user experience will be facilitated by the further involvement of end-users and output checkers.

\section*{ACKNOWLEDGMENT}
The authors are grateful to Elizabeth Green and Felix Ritchie for comments as this work progressed.

\bstctlcite{IEEEexample:BSTcontrol}
\bibliographystyle{IEEEtran}
\bibliography{abrv,references}

\vspace{-0.6\baselineskip}
\begin{IEEEbiography}[{\includegraphics[width=1in,height=1.25in,clip,keepaspectratio]{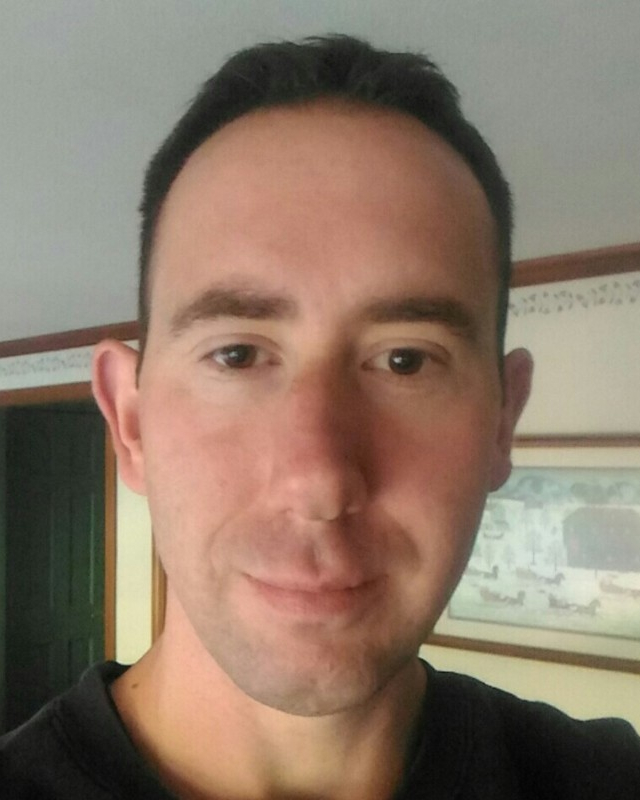}}]
{RICHARD~J.~PREEN}~received the B.Sc.\ (Hons.) and M.Sc.\ degrees in Computer Science in 2004 and 2008, and the Ph.D.\ degree in Artificial Intelligence in 2011, from the University of the West of England (UWE), Bristol, UK\@. He is currently a Senior Research Fellow with the School of Computing and Creative Technologies at UWE\@. His research interests include evolutionary computation, machine learning, surrogate modelling, energy production, and privacy-enhancing technologies.
\end{IEEEbiography}

\vspace{-0.6\baselineskip}
\begin{IEEEbiography}[{\includegraphics[width=1in,height=1.25in,clip,keepaspectratio]{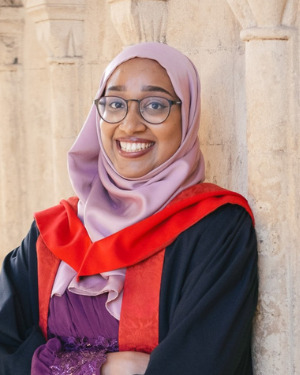}}]
{MAHA~ALBASHIR}~received the B.Sc.\ degree in Electrical and Electronic Engineering, specialising in electronic systems and software engineering, from the University of Khartoum, Khartoum, Sudan, in 2017, and the M.Sc.\ degree in Data Science, from the University of the West of England (UWE), Bristol, UK, in 2022. She was a Research Software Engineer at UWE, developing privacy disclosure assessment tools. She is currently exploring opportunities in data science and machine learning. 
\end{IEEEbiography}

\vspace{-0.6\baselineskip}
\begin{IEEEbiography}[{\includegraphics[width=1in,height=1.25in,clip,keepaspectratio]{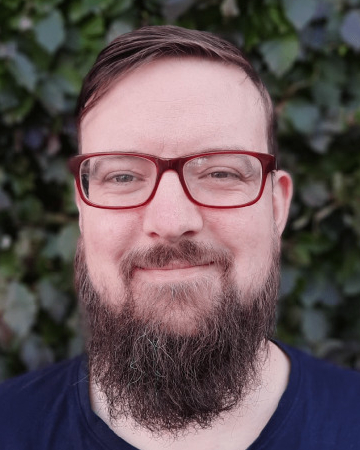}}]
{SIMON~DAVY}~received the B.Sc.\ (Hons.) degree in Computing in 2002, and the Ph.D.\ degree in Artificial Intelligence for Cloud Computing in 2008, from the University of Leeds, UK\@. He is currently a Senior Platform Engineer with the Bennett Institute for Applied Data Science at the University of Oxford, UK\@. His work focuses on the operation and security of the OpenSAFELY platform at the University of Oxford. He was the initial lead developer for the UK National Pathology Exchange.
\end{IEEEbiography}

\vspace{-0.6\baselineskip}
\begin{IEEEbiography}[{\includegraphics[width=1in,height=1.25in,clip,keepaspectratio]{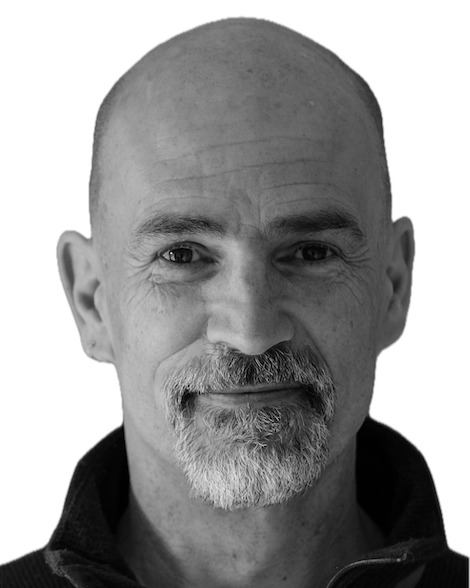}}]
{JIM~SMITH}~received the B.A.\ (Hons.) degree in Electrical Sciences in 1986 from the University of Cambridge, UK, and the Ph.D.\ degree from the University of the West of England (UWE), Bristol, UK, in 1998. He is currently a Professor of Interactive Artificial Intelligence with the School of Computing and Creative Technologies, UWE\@. His  research interests include the intersection of AI and Statistical Disclosure Control; human-machine teaming; and `self-adaptive' intelligent systems that change their learning strategies in response to experience.
\end{IEEEbiography}

\vfill\pagebreak

\end{document}